# Astrobites as a Community-led Model for Education, Science Communication, and Accessibility in Astrophysics

**Submitted as a State of the Profession Consideration to the Astro2020 Decadal Survey**

**Authors:**

Gourav Khullar (gkhullar@uchicago.edu, University of Chicago)
Susanna Kohler (AAS)
Tarini Konchady (Texas A&M University)
Mike Foley (Harvard University)
Amber L. Hornsby (Cardiff University)
Mithi A. de los Reyes (Caltech)
Nora Elisa Chisari (University of Oxford)
V. Ashley Villar (Harvard University)
Kaitlyn Shin (MIT)
Caitlin Doughty (New Mexico State University)
Nora Shipp (University of Chicago)
Joanna Ramasawmy (University of Hertfordshire)
Zephyr Penoyre (Columbia University)
Tim Lichtenberg (University of Oxford)
Kate Storey-Fisher (NYU)
Oliver Hall (University of Birmingham)
Briley Lewis (UCLA)
Aaron B. Pearlman (Caltech)
Alejandro Cárdenas-Avendaño (UIUC)
Joanna S. Bridge (University of Louisville)
Elena González-Egea (University of Hertfordshire)
Vatsal Panwar (University of Amsterdam)
Zachary Slepian (UC Berkeley)
Mara Zimmerman (University of Wyoming)

**Co-signers | Endorsers:**

Emily Sandford, Columbia University
Kelly Malone, Los Alamos National Laboratory
Camille Avestruz, University of Chicago

Jessica May Hislop, Max Planck Institute of Astrophysics
Laura Schifman, Boston University | envirobites.org
Ingrid Pelisoli, University of Potsdam | astropontos.org
Michael Zevin, Northwestern University
Anna Rosen, Center for Astrophysics | Harvard & Smithsonian
Tanveer Karim, Center for Astrophysics | Harvard & Smithsonian
Daniel Palumbo, Center for Astrophysics | Harvard & Smithsonian
Rohan Naidu, Center for Astrophysics | Harvard & Smithsonian
J. Allyn Smith, Austin Peay State University
Tomas Ahumada, University of Maryland
Jeyhan Kartaltepe, Rochester Institute of Technology
Giovanni Fazio, Center for Astrophysics | Harvard & Smithsonian
Ben Cook, Center for Astrophysics | Harvard & Smithsonian
Briley Lewis, University of California, Los Angeles
Stephanie Hamilton, University of Michigan
Charles Law, Center for Astrophysics | Harvard & Smithsonian
Aaron Patton, Pennsylvania State University
Locke, Center for Astrophysics | Harvard & Smithsonian
Munazza Alam, Center for Astrophysics | Harvard & Smithsonian
Mason Ng, Massachusetts Institute of Technology
Mercedes López-Morales, Center for Astrophysics | Harvard & Smithsonian
Bhawna Motwani, Columbia University
Michael B. Deaton, Schweitzer Engineering Laboratory
Jacob White, Konkoly Observatory
Ian Weaver, Center for Astrophysics | Harvard & Smithsonian
Nathan Sanders, Astrobites
Chuanfei Dong, Princeton University
Kerry Hensley, Boston University
Azadeh Keivani, Columbia University
Joshua Pepper, Lehigh University
Emily Gilbert, University of Chicago
Edwin Kite, University of Chicago
Josh Fuchs, Texas Lutheran University
Andrew Chael, Center for Astrophysics | Harvard & Smithsonian
Edward Gomez, Las Cumbres Observatory
John Carlstrom, University of Chicago
Andrew Burkhardt, Center for Astrophysics | Harvard & Smithsonian
Griffin Hosseinzadeh, Center for Astrophysics | Harvard & Smithsonian
Juliana Garcia-Mejia, Center for Astrophysics | Harvard & Smithsonian

Chantanelle Nava, Center for Astrophysics | Harvard & Smithsonian
Ellis Avallone, University of Hawaii at Manoa Institute for Astronomy
Sarah Sadavoy, Center for Astrophysics | Harvard & Smithsonian
Irene Shivaei, University of Arizona
Calvin Leung, MIT
Kedar Phadke, University of Illinois Urbana-Champaign
Samuel Factor, The University of Texas at Austin
Nicholas Young, Michigan State University | PERbites.org
Vivian U, University of California, Irvine
Katya Gozman, University of Chicago
Chris Lintott, University of Oxford/AAS Journals
Brian Nord, Fermilab
Michael Calzadilla, MIT
Shmuel Bialy, Center for Astrophysics | Harvard & Smithsonian
Jennifer Sieben, Indiana University
Mattia Negrello, Cardiff University, UK
Lia Corrales, University of Michigan
Kim Coble, San Francisco State University
Rebecca Diesing, University of Chicago
Rostom Mbarek, University of Chicago
Leah Fulmer, University of Washington
Mary Odekon, Skidmore College
Jason Poh, University of Chicago

**THE ASTRO-PH READER'S DIGEST | SUPPORTED BY THE AAS**

**Abstract:**

Support for early career astronomers who are just beginning to explore astronomy research is imperative to increase retention of diverse practitioners in the field. Since 2010, Astrobites has played an instrumental role in engaging members of the community — particularly undergraduate and graduate students — in research. In this white paper, the Astrobites collaboration outlines our multi-faceted online education platform that both eases the transition into astronomy research and promotes inclusive professional development opportunities. We additionally offer recommendations for how the astronomy community can reduce barriers to entry to astronomy research in the coming decade.

**Outline:**

---

# 1. Introduction

Making the astronomy community more inclusive, diverse and accessible is an ongoing effort. In recent years, a number of national task forces have explored ways to improve astronomy graduate education in response to a changing culture ([1], [2]). These reports' findings underscore the need to develop tools to better prepare students for professional careers in astronomy after their undergraduate studies. As most professional astronomers conduct research, undergraduates are expected to gain some experience with the body of astronomy research literature as well as the process of scientific research. But the steep learning curve associated with this undertaking means that lack of support and assistance for undergraduates at this stage can act as an entry barrier to those otherwise interested in pursuing careers in astronomy.

Multiple approaches have been taken to help ease entry to astronomy research. A number of universities offer Research Experience for Undergraduate (REU) programs in astronomy [3], which allow undergraduate students to gain research experience before they apply for graduate programs. However, REUs are only reaching a small and specific community of physics and astronomy students worldwide. There are currently 26 active astronomy REU sites in the US, typically serving cohorts of 8–10 astronomy and physics undergraduates per year. In 2017, 535 students graduated with a bachelor's degree in astronomy in the US; with ~10% increase in the astronomy student population every year [4]. Other similar efforts include summer research programs offered specifically for underrepresented minority (URM) students [5] and "bridge programs" [6]; once again, such programs serve only a small fraction of the undergraduate population and cannot serve all URM undergraduate students worldwide. Many graduate and undergraduate programs hold journal clubs for the purpose of discussing recent research papers. While these journal clubs provide a valuable opportunity for earlier-stage students to ask questions and present material, there is still a significant leap from undergraduate research and academics to parsing and discussing active research.

*Lowering the structural barriers to entry that restrict access to astronomy research is imperative, especially for minoritized groups.* The limited scope of the many aforementioned efforts to improve access to astronomy research means that the astronomy community can benefit from investment in alternative programs that reach large, diverse audiences. Online platforms like Astrobites are excellent examples of such programs. Through Astrobites, we aim to provide a multi-faceted online education platform that both eases the transition into astronomy research and promotes inclusive professional development opportunities.

## 2. Description of Astrobites

Astrobites is an American Astronomical Society (AAS)-supported daily astrophysical literature journal written by graduate student volunteers. At its creation in 2010, the site's intent was to present one astronomy research paper per day in a brief, accessible format for undergraduate students interested in active research in astronomy. Since its inception, Astrobites has grown into not just an archive of accessible summaries of recent astronomy research, but also a host of other resources for undergraduate students and the broader community (see Figure 1 for an overview).

## 2.1 Content

Astrobites posts are typically written by a rotating team of volunteer graduate student authors recruited each fall (see Figure 3). The site also regularly publishes guest posts submitted by graduate students outside of Astrobites's normal authorship, as well as by astronomy faculty, researchers, and even undergraduate students. Guest authors work with Astrobites editors to prepare their posts, providing opportunities for students to gain writing and editing experience and increasing the diversity of voices seen on Astrobites.

### 2.1.1 Daily Paper Summaries

Astrobites, at its core, is a reader's digest of current astrophysical literature. The standard post is a *Daily Paper Summary*, in which an Astrobites author briefly summarizes a recent article published on the arXiv, a scientific research preprint server. Daily Paper Summaries are typically under 1,000 words and describe not just the key results from the selected research study, but also the relevant background and context and the core methods to obtain their results. Posts include reproductions and explanations of key figures from the study and link back to the original paper, and Astrobites authors work to illustrate the process of science as much as the outcomes of an individual study. Over its lifetime, Astrobites has built an archive of nearly 2,000 Daily Paper Summaries.

Although most posts summarize research published within the last few months, Astrobites authors also sometimes highlight *Astrophysical Classics* — historical research papers that the community now recognizes as seminal works that advance the field. By exploring these classics, Astrobites hopes to provide helpful context for new researchers who may not yet be aware of the past discoveries that are driving current research questions.

### 2.1.2 Conference Coverage

Astrobites has been reporting on bi-annual AAS and European Week of Astronomy and Space Science (EWASS) meetings since 2012. *Conference Coverage* posts include a series of interviews conducted by Astrobiters of plenary speakers and award winners at each conference, as well as daily live-blogging of a broad selection of conference sessions. The interviews give our readers access to the speakers' research and career paths, as well as their talks. Live-blogging coverage allows members of the astronomy community to follow and recap specific sessions that they were unable to attend. Astrobites coverage of meetings also provides an introduction that may make future conference attendance less intimidating to early-career readers, and the interviews with plenary speakers help to make these high-profile researchers more relatable and increase the accessibility of the field.

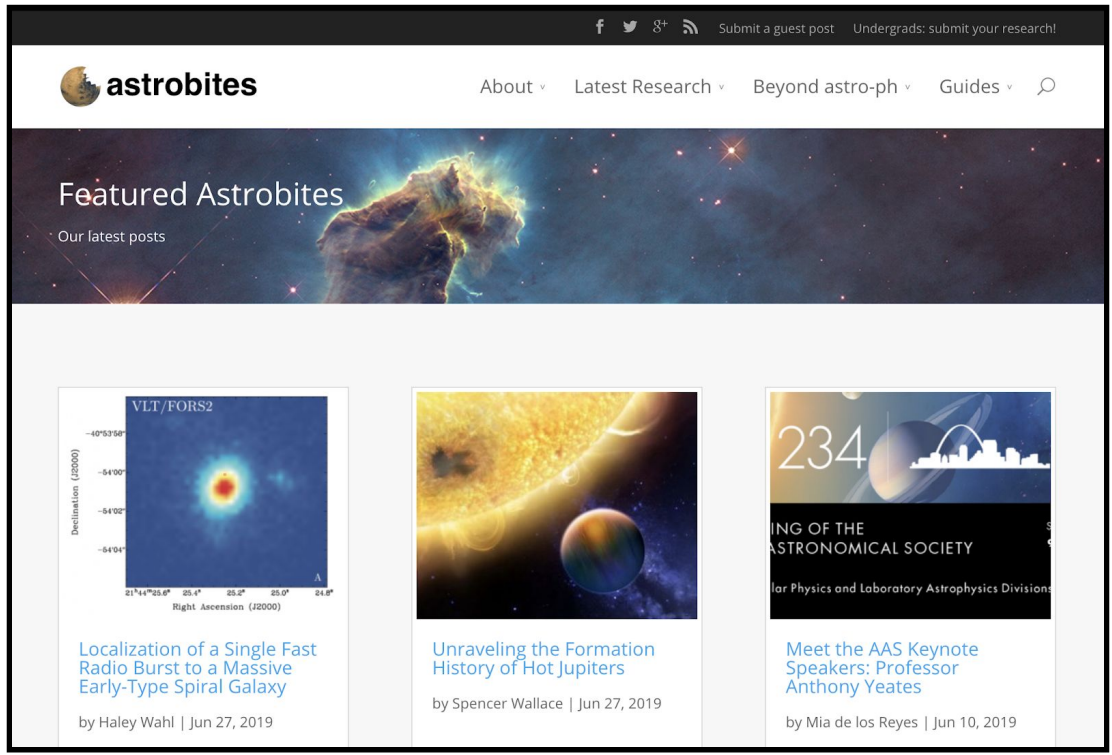

**Figure 1.** Screenshot of the Astrobites website on June 29th, 2019, with posts on arXiv papers on fast radio bursts and hot Jupiters, as well as an interview with a keynote speaker at the 234th AAS Meeting. A call for guest posts, including from undergraduate researchers, and tabs for "Beyond astro-ph" posts and "Guides" (see Sec 2.1.3) are also visible.

### 2.1.3 Beyond Astro-ph

Beyond the daily summaries of recent or classic papers in astrophysics, Astrobites regularly publishes pieces on other topics relevant to early career astronomers by taking a multi-faceted approach. These *Beyond astro-ph* pieces, published approximately once every two weeks, touch on the following topics:

- **Career navigation & personal experiences:** These posts provide perspectives from various members of the astronomical community, describing their career paths and experiences. The main goal of these posts is to introduce students to a wide range of astronomy-related careers, both in and outside academia. Posts like these can also help junior astronomers to get to know other members in the community.

- **Resources:** These posts broadly provide resources for junior researchers. Some of these posts describe research-related topics and help readers gain new skills (e.g. basic techniques for working with astrophysical simulations and data), while others explain how to apply for REUs, graduate schools, and fellowships. The goal is to provide information for budding astronomers, particularly those who may have limited access to resources at their institutions.

- **Diversity, equity, and inclusion (DEI):** These posts highlight studies on DEI that affect the astronomical community. Recent DEI posts have discussed topics like LGBTQIA+ inclusion in the field, inclusion-themed AAS workshops, and the report on sexual harassment by the National Academies of Sciences, Engineering, and Medicine. Topics of inclusion and social justice are a major part of the current cultural discourse in astronomy. By participating in that conversation, Astrobites strives to support minoritized astronomers in the field, and to increase awareness that astronomy is fundamentally a human endeavor and therefore subject to societal issues.

- **Science policy:** These posts discuss topics related to science policy—both how policy and politics can affect astronomers (posts often discuss current events, such as Brexit, and the ongoing Thirty Meter Telescope construction conflict), and how astronomers can affect policy (emphasizing how students can get involved). Astrobites liaises with the AAS Bahcall Policy Fellow, who helps suggest post topics and fact-checks drafts. The goal is to highlight the importance of policy to astronomy, a topic not often discussed in the average astronomy classroom, yet is of utmost importance to the development of international collaborations and productive long-term national research programs.

## 2.2 Astrobites in the Classroom

Astrobites is increasingly playing a vital role in classrooms by enhancing access to scientific publications. Several graduate and undergraduate programs have incorporated Astrobites in their curricula, and the Astrobites collaboration has assembled lesson plans to facilitate the use of Astrobites in the classroom.

Specific recommendations on how to incorporate Astrobites into astronomy and physics coursework are given in Sanders et al. 2017 [7]. This article is the result of feedback from a number of astronomy instructors who have successfully integrated Astrobites into their classrooms, and it details a variety of pedagogical tools to enable astrophysicists to teach with Astrobites, including three sample lesson plans and grading rubrics. The lesson plans suggest uses of Astrobites in the classroom at various educational stages (early undergraduate through graduate).

In addition to publishing educational resources, the Astrobites collaboration is building on earlier work by conducting a more detailed study on Astrobites's efficacy as an educational tool. Supported by the AAS Education & Professional Development (EPD) Mini-Grant Program from 2017–2019, we conducted education workshops at the 231st AAS Meeting 2018 and the EWASS meeting in 2019 on using Astrobites as an objective-based, student-focused teaching tool in undergraduate and early graduate astronomy courses. We also invited educators to participate in our research study via a focus group and collected data on their experiences using Astrobites in their classrooms. The lesson plans can be accessed here; educators can participate in the ongoing study via this form.

## 2.3 Expansion of the Astrobites Model

Since 2010, Astrobites has expanded into a network of "Science Bites" websites led by graduate students and postdocs in different fields and in different languages. At the moment, The Science Bites family encompasses 17 sites and covers 15 different scientific topics, among which there are several focused in physics and astrophysics: soft matter (Softbites), particle physics (particlebites), physics education research (PERbites), and astroparticle physics (astroparticlebites). Additional sites have arisen dedicated to chemistry (Chembites), oceanography (Oceanbites), cognitive science (Cogbites), oncology (Oncobites), immunology (immunobites), environmental science (Envirobites), evolutionary biology (evobites), and coral reef science (reefbites).

Furthermore, there are several sites that feature content in other languages, among them: Astrobites in Spanish (astrobitos), Portuguese (astropontos), and Persian/Farsi (staryab) (see Figure 2). Oceanbites and Envirobites also feature content in Spanish, and sites such as Astrobitos have started cross-posting material from Astrobites, thus promoting interdisciplinarity in general as well as fostering further interaction between site administrators and writers. Some of these sites, such as staryab, arose independently of Astrobites and later became part of the Science Bites network. The advent of such an extended network of sites in such a short time has demonstrated that the Astrobites model has successfully filled a gap in how researchers communicate with the broader community. Posts in different languages have broadened the community, increased its diversity, and aim to reduce the divide in resources caused by the lack of critical mass of the science community and the language barrier in certain regions of the world.

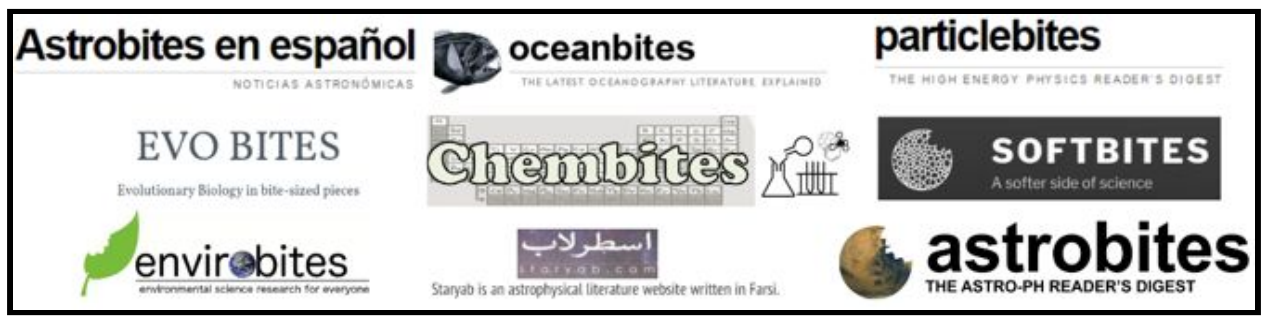

**Figure 2.** The Astrobites model has grown into the ScienceBites network across various STEM fields, a few of which are shown in this figure. See Section 2.3 for further detail.

## 2.4 Readership

Primarily via word of mouth, Astrobites's readership has built up to nearly 46,000 pageviews each month from over 15,000 users in 2018 (see Figure 4). Readership spans all populated continents; just over 50% of Astrobites's readers come from the United States, and the next most highly represented countries are the United Kingdom, India, Canada, and Germany.

Surprisingly, regular surveys of the Astrobites readership reveals that only a small fraction of regular readers (10–20%) are within our focus audience of undergraduate students. The bulk of the site's readership is made up of graduate students (30–40%), researchers (10–30%), and astronomy enthusiasts (15–30%). This unexpected boost in readership from the broader astronomy community presumably demonstrates the unmet need that the site addresses for brief and accessible summaries of recent astronomy research.

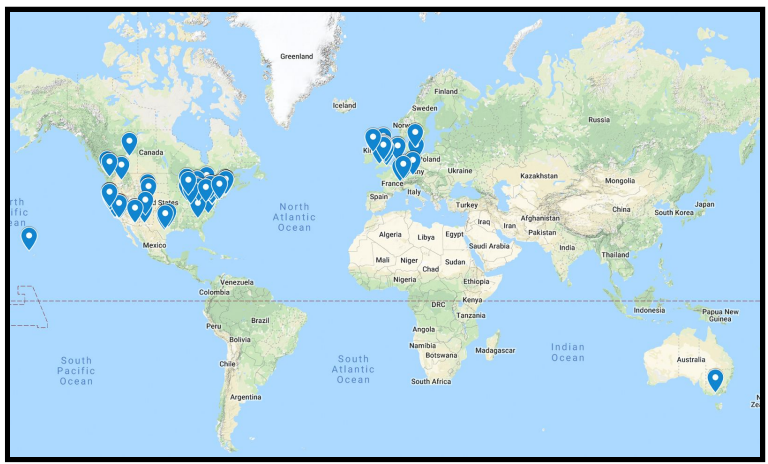

**Figure 3.** Location information for our current authors and alumni. We do not currently have any authors outside North America, Europe, and Australia; active hiring from and outreach to departments worldwide is a major ongoing project.

## 3. Broader Impacts of Astrobites

In addition to serving undergraduate students, Astrobites impacts the broader astronomy community and beyond:

- **Professional astronomers:** In a time when progress and advancement in the field is often measured via publication record [8], Astrobites's role in building publication-parsing skills in early career astronomers is vital. However, the rate at which new papers are submitted to the arXiv ensures that even trained professional astronomers may struggle to remain up to date with the broader field. The concentrated nature of Astrobites summaries benefits professional astronomers by providing exposure to significant research results that may otherwise slip under their radar. Additionally, Astrobites serves as an additional press outlet to highlight recent work of professional researchers, via our website, social-media reach, and trackback links to the arXiv.

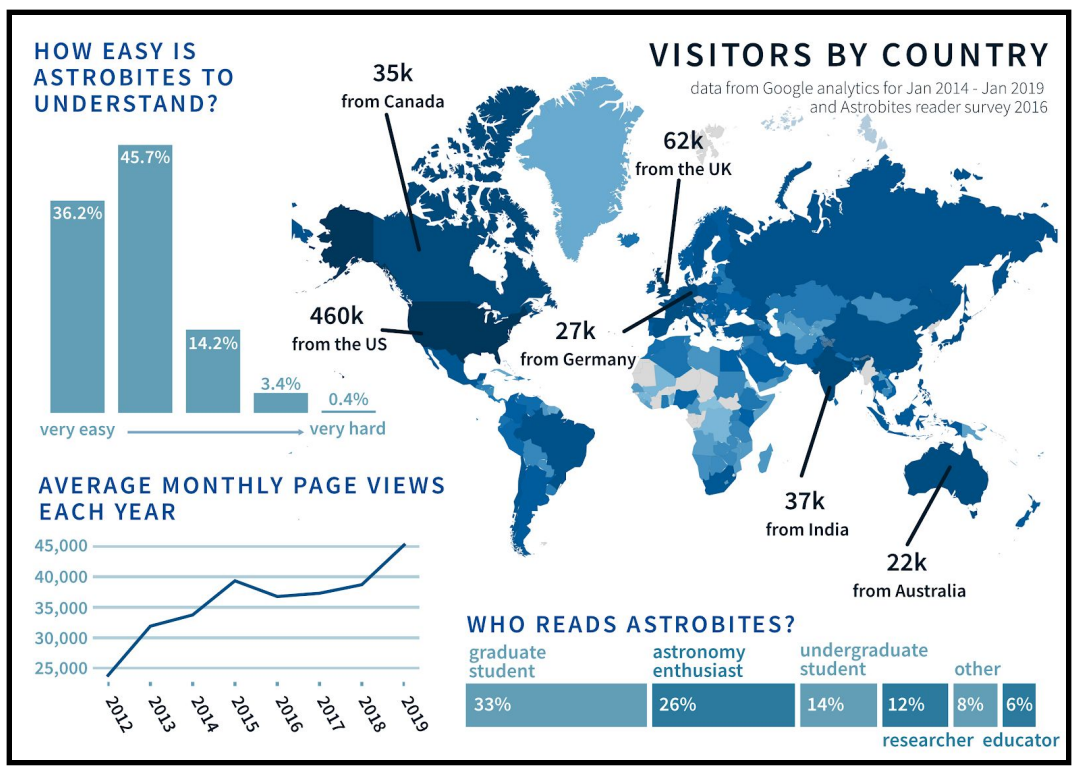

**Figure 4.** Top left and bottom right: Results and highlights from a 2016 readership survey. Top right and bottom left: Readership statistics by country and by year, from Google analytics (2014–2019).

- **Astrobites graduate-student writers:** Astrobites authors, many of whom are first- or second-year graduate students, must study scientific papers in great detail in order to thoroughly and accurately relate the salient aspects of the work to the Astrobites readership. Not only does this type of writing develop the skills of efficiently reading scientific papers and mentally highlighting the crucial aspects of the work, this also trains authors in science communication ([9]–[12]); indeed, several Astrobites alumni have pursued successful careers in science writing and communication.

  Additionally, through writing articles and interviewing conference keynote speakers, Astrobites writers can become familiar with and gain introductions to leaders of the field early in their scientific careers. Authors also have access to the full network of over 100 Astrobites alumni, many of whom are now astronomy and physics faculty.

  Finally, Astrobites members gain crucial management skills by serving as administrators for the collaboration: authors sit on or chair committees (e.g. on matters of policy, diversity, equity and inclusion, education research, scheduling, etc.), organize and run meetings, and give productive feedback to others.

- **Benefits to the astronomy community:** Our work at Astrobites benefits minoritized members of the community. Along with reducing gatekeeping and increasing access to publications, Astrobites reports on inclusion efforts across astronomy, highlights issues of marginalization faced by astronomers, and is in the process of compiling resources for peer-education and training in conversations surrounding diversity, equity, and inclusion (DEI). We also actively work on outreach efforts in countries with growing astronomy communities, both towards readers and new authors.

- **Benefits to society:** It is more important than ever that the non-expert public have access to accurate science reporting. Although Astrobites's intended audience is undergraduates studying astronomy/physics, the non-expert public is a significant segment of the site's current readership. Astrobites strives to reduce scientific jargon and clearly explain the principles behind the research, making recent scientific studies accessible to each reader.

## 4. Recommendations

We strongly believe that online platforms like Astrobites will serve as important educational and community tools in the upcoming decade. To fully realize our vision for such platforms, we make the following recommendations for the broader astronomical community. These recommendations are based on our experience in creating and maintaining Astrobites and our research on its use in astronomy classrooms.

- **Use resources like Astrobites in the classroom**
  Although reading, parsing and analysing the scientific literature is an essential skill for astronomers, it is rarely taught in undergraduate programs. Platforms like Astrobites allow students to become familiar with the literature and to practice scientific communication early in their academic careers. Our education research study and lesson plans are resources that facilitate this. Moreover, financially supporting structures like the AAS Education & Professional Development (EPD) Mini-Grant Program are concrete steps that enable astronomy pedagogy research.

- **Promote platforms like Astrobites to reduce gatekeeping and increase diversity**
  Astrobites posts provide a much-needed resource that allows undergraduate students, early graduate students, and the broader community to gain access to astronomy research. As mentioned in Section 2.1.3, our posts also boost conversations on social issues in the field; promoting these multifaceted and broad initiatives is an essential step towards reducing entry barriers for learning and training, especially for minoritized astronomers.

- **Develop Bites-based journal clubs**
  Journal clubs are a typical resource in the professional community, providing an environment for colleagues to learn about, discuss, and critique the literature. While undergraduates are currently expected to "learn by osmosis", resources like Astrobites provide accessible entry points for undergraduates to actively join in these discussions. We encourage departments to form journal clubs that are specifically designed for undergraduates and early graduate students to create a non-intimidating environment in which to learn how to read papers, become familiar with the literature, and provide constructive criticism to the papers they read.

- **Sponsor research on the efficacy of Bites programs**
  Although some preliminary research exists ([7]), we still need a thorough study that evaluates the impacts that introduction-to-research programs like Astrobites have on early career astronomers, and how Bites programs (and similar initiatives) can improve to best suit the educational and career needs of their audience.

- **Increase financial and structural support for Bites programs**
  Currently our Bites programs, including Astrobites, are run entirely by community volunteers and operate on a lean budget, with server support and funding for promotional material like stickers and posters generously provided by the American Astronomical Society. Some of the sites of the network, like [staryab](#), are self-funded. Increased community financial investment in grassroots efforts like the Bites sites, however, could significantly improve the efficacy and reach of these programs. As an example, with a pool of external funding, Astrobites could afford to cover travel costs for authors to attend national and international astronomy conferences, where they can engage in live-blogging, community-building, and facilitating education workshops. Such external support would remove the barrier for volunteers who wish to help expand the Astrobites program but cannot provide their own travel funding.

- **Increase support for graduate-student education in scientific communication**
  Scientific communication is an essential skill for all scientists [13]. Bites programs provide a concrete method to provide training to graduate students in scientific communication at no cost to individual universities. However, summarizing scientific literature for undergraduates is only one skill within the toolbox of scientific communication. Financial support for students to attend science-communication conferences [e.g., 14], workshops, and other professional development opportunities will expose them to the full toolset available and further train them to become better communicators — and, ultimately, scientists.

In the coming decade, we believe that the multifaceted and broad approach of initiatives like Astrobites is essential to reduce barriers to entry into astronomy research. Through this framework, we strive to smooth the way for students seeking careers in astronomy and to build a more inclusive and accessible culture in the field.